\documentclass{revtex4}

\usepackage{amsthm,amssymb,amsmath}				
\usepackage{graphicx,comment}
\usepackage{float}   
 \usepackage{xcolor}

\begin{document}

\title{KG-oscillators in Eddington-inspired Born-Infeld gravity: Wu-Yang magnetic monopole and Ricci scalar curvature effects.}
\author{Omar Mustafa}
\email{omar.mustafa@emu.edu.tr}
\affiliation{Department of Physics, Eastern Mediterranean University, 99628, G. Magusa,
north Cyprus, Mersin 10 - Turkiye.}

\begin{abstract}
\textbf{Abstract: }We investigate the Klein-Gordon (KG) oscillators in a global monopole (GM) spacetime in Eddington-inspired Born-Infeld (EiBI) gravity and a Wu-Yang magnetic monopole (WYMM). We discuss the gravitational effects in the presence of  Ricci scalar curvature $R=R_{\upsilon }^{\upsilon }$. It is observed that the presence of the Ricci scalar curvature, effectively and manifestly, introduces a force field that makes the corresponding quantum mechanical repulsive core more repulsive. Similar effect is also observed for the EiBI-gravitational field.  We reiterate and report that the corresponding bosonic KG-oscillator quantum mechanical system admits a solution in the form of confluent Heun functions, the truncation of which into a physically admissible polynomial is shown to be associated with some parametric correlations/conditions. The use of such conditions/correlations is mandatory and yields a set of allowed/restricted quantum mechanical orbital $\ell $-excitations, for all radial quantum numbers $n_{r}\geq 0$. Our procedure is shown to be quite handy, in the sense that it allows one to retrieve results for KG-oscillators in GM-spacetime in different EiBI-gravity and Ricci scalar curvature settings.


\end{abstract}

\maketitle

\section{Introduction}

The rapid expansion and cooling process of the early universe is believed to 
feasibly generate some topological defects in the spacetime fabric \cite%
{R1,R2}, like domain walls \cite{R020,R021}, cosmic strings \cite%
{R021,R3,R4,R5,R6} and global monopoles (GM) \cite{R2,R3,R6,R7}. Global
monopoles are spherically symmetric topological defects, that effectively
exert no gravitational force, and the space around and outside such
monopoles has a deficit angle that reflects all light, e.g., \cite{R2}.
Their line element metric is given by%
\begin{equation}
ds^{2}=-dt^{2}+\frac{1}{\alpha ^{2}}dr^{2}+r^{2}\left( d\theta ^{2}+\sin
^{2}\theta \,d\varphi ^{2}\right) ,  \label{In1}
\end{equation}%
where $0<\alpha ^{2}=1-\tau \leq 1$, $\tau =8\pi G\zeta ^{2}$ is the deficit
angle, $\alpha $ is a GM parameter that takes into account the energy scale $\zeta$, and $G$ is the gravitational constant \cite%
{R7,Re8,Re9,Re91,Re10}. Obviously, this metric collapses into the flat
Minkowski one when $\alpha =1$. Over the years, such a metric has stimulated
intensive research attention to study, for example, the spacetime geometry
around GM within $f\left( R\right) $ theories of gravity \cite{Re100},
the vacuum polarization effects in the presence of a Wu-Yang magnetic
monopole (WYMM) \cite{Re101,Re1011,Re102}, and gravitating magnetic monopole
\cite{Re103,Re104}. Moreover, the effects of GM spacetime on the
spectroscopic structure of some quantum mechanical systems are studied.
Amongst are, Dirac and Klein-Gordon (KG) oscillators \cite{Re11}, Schr\"{o}dinger oscillators \cite{Re9}, 
Schr\"{o}dinger oscillators in a GM spacetime
and a Wu-Yang magnetic monopole \cite{Re91}, KG particles with a dyon,
magnetic flux and scalar potential \cite{Re8}, bosons in Aharonov-Bohm (AB)
flux field and a Coulomb potential \cite{Re131}, Schr\"{o}dinger particles
in a Kratzer potential \cite{Re132}, Schr\"{o}dinger particles in a Hulth%
\.{e}n potential \cite{Re1321} , scattering by a monopole \cite{Re133}, Schr%
\"{o}dinger particles in a Hulth\.{e}n plus Kratzer potential \cite{Re1331},
KG-oscillators and AB-effect \cite{Re1332}. Yet, the influence of
topological defects associated with different spacetime backgrounds on the
quantum mechanical systems has been a subject of
research attention over the years. Like Dirac and Klein-Gordon (KG)
oscillators in a variety spacetime structures, e.g., \cite%
{Re11,Re12,Re121,Re13,Re14,Re15,Re16,Re17,Re18,Re19,Re20,Re21,Re211,Re212,Re213,Re22,Re23,Re24,Re25,Re26,Re27,Re271,Re272,Re273}.

However, the inclusion of Born-Infeld nonlinear electrodynamics into Eddington theory of gravity \cite{ref1,ref11,ref12} has inspired the notion \textit{ Eddington-inspired Born-Infeld} (EiBI) theory of gravity. This is, in fact, an equivalent theory to Einstein's General Relativity (GR) in vacuum. EiBI has, nevertheless, additional distinctive features (when matter
is included) and possesses internal consistency (in the sense that it is
free of instabilities and ghosts \cite{ref2}). EiBI-gravity yields entirely
cosmological singularity-free universe \cite{ref1,ref3} (which is its most
intriguing feature), and may feasibly accommodate compact stars \cite%
{ref31,ref32,ref33}. Therefore, different relevant research studies are carried out. For example, the study of the cosmological consequences of Born-Infeld gravity \cite{ref34}, of Born-Infeld determinantal gravity and the taming of the conical singularity in 3-dimensional spacetime \cite{ref35}, of the Born-Infeld extension of new massive gravity \cite{ref36},  of the unitarity analysis of general Born-Infeld gravity theories \cite{ref37}, and a comprehensive review on alternative theories of gravity \cite{ref38}, to mention a few.

On the other hand, the GM spacetime in EiBI-gravity, generated by a source
matter, is described \cite{ref3,ref4,ref5,ref6}, in spherical coordinates,
by the metric%
\begin{equation}
ds^{2}=-\alpha ^{2}\,dt^{2}+\frac{r^{2}}{\alpha ^{2}\left( r^{2}+\kappa \tau
\right) }\,dr^{2}+r^{2}\,\left( d\theta ^{2}+\sin ^{2}\theta \,d\varphi
^{2}\right) ,  \label{In2}
\end{equation}%
where $\kappa $ the Eddington parameter. One should notice that for $\kappa =0$ (i.e.,\ no
Eddington gravity), such a spacetime collapses into that of GM in (\ref{In1}%
) (hence, occasionally called EiBI-GM). The study of the gravitational field
effects of EiBI-gravity on the quantum mechanical spectroscopic structure
has only very recently been carried out by Pereira et al \cite{ref5,ref7}.
Therein, the authors have studied Klein-Gordon (KG) oscillators in GM
spacetime in EiBI-gravity. They were able to report their results for only $%
n=1$ state ( for any value of the angular momentum quantum number $\ell $). The methodology they have followed, although good,  does not work for $n>1$ states. We have, however, introduced an alternative approach \cite{ref701} to obtain a \emph{conditionally exact solution} (or if you wish, \emph{quasi-exact solution}) \cite{ref8} for the Klein-Gordon (KG) oscillator in a GM-spacetime within EiBI-gravity and in a WYMM. Such an alternative approach works for any $n$ and $\ell $ states, provided that the KG-oscillator's frequency and the Eddington parameter are conditionally correlated. To the best of our knowledge, the studies by Pereira et al \cite{ref5,ref7} and Mustafa et al.  \cite{ref701} are the only attempts made, in the literature, in this regard. 

We believe that a comprehensive understanding of the effects of gravitational fields, on the spectroscopic structure of elementary particles, in different spacetimes becomes more accurate and more meticulous when a wider set of quantum states with $n\geq 1$ is considered. This fact, along with just the few studies (mentioned above) on the spectroscopic structure of the KG-oscillators in EiBI-gravity (including the effect of the Ricci scalar curvature, $R=R_{\upsilon }^{\upsilon }$ field, and a WYMM) form our motivation in the current methodical proposal. The current study shall not only partially fill this gap but also be of fundamental pedagogical interest for quantum gravity and condensed matter physics (e.g., \cite{ref9,ref10,ref011}).

The organization of our study is in order. In Section 2, we discuss KG-oscillators
in a GM spacetime within EiBI-gravity in a WYMM, including the Ricci scalar curvature $R=R_{\upsilon }^{\upsilon }$ effect. We observe that the presence of the Ricci scalar curvature makes the corresponding quantum mechanical repulsive core, $\mathcal{\ell }\left( \ell +1\right)
/r^{2}\Longrightarrow \mathcal{\tilde{L}}\left( \mathcal{\tilde{L}+}1\right)
/r^{2}$, more stronger, whereas the WYMM makes it weaker. Moreover, we bring the radial part of such a quantum mechanical system into a form that admits confluent Heun functions $%
H_{C}\left( \alpha ,\beta ,\gamma ,\delta ,\eta ,z\right) $ solution. The truncation of which into a polynomial of order $n_{r}+1;\,n_{r}\geq 0$, is secured by the condition $\delta =-\alpha \left( n_{r}+\left[ \beta
+\gamma +2\right] /2\right) $ (cf., e.g., \cite{ref701,ref71}). In Section 3, we discuss one of the feasible conditions/correlations (in addition to those reported by Ishkhanyan et al. \cite{ref71}) that allows such a truncation recipe to be valid. It is in our opinion, therefore, that the use of such conditions/correlations is mandatory. Consequently, this truncation condition would result in a set of allowed quantum-mechanical orbital $\ell $-excitations. This is discussed and reported in Section 3. The same procedure is shown to be quite flexible and handy, in the sense that it allows one to obtain results for KG-oscillators in GM-spacetime in EiBI-gravity (i.e., $\kappa \neq0$) with Ricci scalar curvature effects, in Subsection 3-A, and the results for KG-oscillators in GM-spacetime in no EiBI-gravity, $\tilde{\kappa}=0$, with the Ricci scalar effects, in Subsection 3-B. We conclude in Section 4.

\section{KG-oscillators in a GM spacetime within EiBI- gravity and in a WYMM including the curvature scalar effects}

We start with a rescaling of time in the forms of $\sqrt{\left( 1-\tau
\right) }\,dt\rightarrow dt$, along with $\tilde{\kappa}=\kappa \tau $, that
allows us to rewrite metric (\ref{In2}) as
\begin{equation}
ds^{2}=-\,dt^{2}+\frac{1}{\alpha ^{2}\left( 1+\frac{\tilde{\kappa}}{r^{2}}%
\right) }\,dr^{2}+r^{2}\,\,\left( d\theta ^{2}+\sin ^{2}\theta \,d\varphi
^{2}\right) .  \label{e1}
\end{equation}%
Hereby, it should be noted that $\tilde{\kappa}<0$ would describe a topologically charged wormhole \cite{ref012,ref013,ref014}, $\alpha =1$ and $%
\tilde{\kappa}<0$ would correspond to a Morris-Thorne-type wormhole
spacetime \cite{ref015,ref016}, $\kappa =0$ would describe a GM spacetime, and $\kappa >0$ corresponds to a GM spacetime in EiBI-gravity. The corresponding Ricci scalar curvature $R=R_{\upsilon }^{\upsilon }$ takes the form%
\begin{equation}
R=2\alpha ^{2}\left[ \frac{\tilde{\kappa}}{r^{4}}+\frac{\left( 1-\alpha
^{2}\right) }{\alpha ^{2}r^{2}}\right].  \label{e50}
\end{equation}%
In this case, the KG-oscillators in EiBI-gravity spacetime, along with a WYMM and a Ricci scalar curvature, would read%
\begin{equation}
\left( \frac{1}{\sqrt{-g}}\tilde{D}^{+}_{\mu }\sqrt{-g}g^{\mu \nu }\tilde{D}^{-}%
_{\nu }\right) \,\Psi \left( t,r,\theta ,\varphi \right) =\left[ m_{\circ
}^{2}+\xi R\right] \,\Psi \left( t,r,\theta ,\varphi \right) ,  \label{e51}
\end{equation}%
where $\tilde{D}^{\pm}_{\mu }=D_{\mu }\pm\mathcal{F}_{\mu }$ ; $\mathcal{F}_{\mu }$ $\in 
\mathbb{R}
$, $D_{\mu }=\partial _{\mu }-ieA_{\mu }$ is the gauge-covariant derivative,  $A_{\nu}=\left( 0,0,0,A_{\varphi }\right) $, and $m_{\circ }$ is the rest mass energy (i.e., $m_{\circ }\equiv m_{\circ }c^{2}$, with $%
\hbar =c=1$ units to be used throughout this study). We may use $\mathcal{F}%
_{\mu }=\left( 0,\mathcal{F}_{r},0,0\right) ;\,\mathcal{F}_{r}=\Omega r$, to incorporate the KG-oscillators in the process. Moreover, the KG-equation is conformally invariant if and only if the coupling constant $\xi$ is given by $\xi =\frac{N-2}{4(N-1)}$, where $N$ is the dimension of the spacetime under consideration, and hence our $N=4 \rightarrow \xi=1/6$ will be used throughout. Under such settings, equation (\ref{e51}) would read%
\begin{gather}
\left\{ -\partial _{t}^{2}+\frac{\alpha ^{2}}{r^{2}}\sqrt{1+\frac{\tilde{%
\kappa}}{r^{2}}}\,\left( \partial _{r}+\mathcal{F}_{r}\right) \,r^{2}\,\sqrt{%
1+\frac{\tilde{\kappa}}{r^{2}}}\left( \partial _{r}-\mathcal{F}_{r}\right) +%
\frac{1}{r^{2}}\left[ \frac{1}{\sin \theta }\,\partial _{\theta }\,\sin
\theta \,\partial _{\theta }\right. \right.    \notag \\
\left. \left. +\frac{1}{\sin ^{2}\theta }\left( \partial _{\varphi
}-ieA_{\varphi }\right) ^{2}\,\right] \right\} \Psi \left( t,r,\theta
,\varphi \right) =\left[ m_{\circ }^{2}+\xi R\right] \,\Psi \left(
t,r,\theta ,\varphi \right) .  \label{e52}
\end{gather}%
To facilitate the separation of variables, we use $\Psi \left( t,r,\theta ,\varphi \right) =e^{-iEt}\phi \left( r\right) Y_{\sigma ,\ell ,m}\left( \theta ,\varphi \right) $, to obtain%
\begin{equation}
\left\{ E^{2}+\frac{\alpha ^{2}}{r^{2}}\sqrt{1+\frac{\tilde{\kappa}}{r^{2}}}%
\,\left( \partial _{r}+\mathcal{F}_{r}\right) \,r^{2}\,\sqrt{1+\frac{\tilde{%
\kappa}}{r^{2}}}\left( \partial _{r}-\mathcal{F}_{r}\right) -\frac{\lambda }{%
r^{2}}\right\} \phi \left( r\right) =\left[ m_{\circ }^{2}+\xi R\right] \phi
\left( r\right) ,  \label{e54}
\end{equation}%
where $\lambda =\ell \left( \ell +1\right) $ for the case $A_{\varphi }=0$. However,  $A_{\varphi }=sg-g\cos \theta ;\,s=\pm 1$, for the WYMM \cite{Re101,Re1011,Re102,ref701} and $\lambda $ is, therefore, given by 
\begin{equation}
\lambda =\ell \left( \ell +1\right) -\sigma ^{2}.  \label{e56}
\end{equation}%
Here $\sigma =eg$, and $g$ is the WYMM strength. More details on the result in Eq.(\ref{e56}) are given in \cite{ref701,ref702}. 

We may now incorporate the KG-oscillator by using $\mathcal{F}%
_{r}=\Omega r$ in (\ref{e54}) to one obtains, in a straightforward manner, 
\begin{equation}
\left\{ \left( 1+\frac{\tilde{\kappa}}{r^{2}}\right) \partial
_{r}^{2}\,+\left( \frac{2}{r}+\frac{\tilde{\kappa}}{r^{3}}\right) \partial
_{r}-\left[ \Omega ^{2}r^{2}+\frac{2\xi \tilde{\kappa}}{r^{4}}\right] -\frac{%
\mathcal{\tilde{L}}\left( \mathcal{\tilde{L}+}1\right) }{r^{2}}+\tilde{E}%
^{2}\right\} \phi \left( r\right) =0.  \label{e57}
\end{equation}%
This equation represents KG-oscillators in GM spacetime in EiBI-gravity and a WYMM along with Ricci scalar curvature, where%
\begin{equation}
\tilde{E}^{2}=\frac{E^{2}-m_{\circ }^{2}}{\alpha ^{2}}-3\Omega -\Omega ^{2}%
\tilde{\kappa},\;\mathcal{\tilde{L}}\left( \mathcal{\tilde{L}+}1\right) =%
\frac{\ell \left( \ell +1\right) -\sigma ^{2}+2\xi \left( 1-\alpha
^{2}\right) }{\alpha ^{2}}+2\Omega \,\tilde{\kappa}.  \label{e58}
\end{equation}%
and hence 
\begin{equation}
\mathcal{\tilde{L}}=-\frac{1}{2}+\sqrt{\frac{1}{4}+\frac{\ell \left( \ell
+1\right) -\sigma ^{2}+2\xi \left( 1-\alpha ^{2}\right) }{\alpha ^{2}}%
-2\Omega \,\tilde{\kappa}}.  \label{e59}
\end{equation}%
This result is in exact agreement with equations (24) and (25) reported by Pereira et al. \cite{ref5} (where our $\Omega =m\omega $ of Pereira et al. \cite{ref5} for $\sigma =0$ in the absence of the WYMM). In the current methodical proposal, however, we include a WYMM and the curvature scalar. Nevertheless, one should observe that the presence of the Ricci scalar curvature, as well as EiBI-gravity, make the quantum mechanical repulsive core, $\mathcal{\tilde{L}}\left( \mathcal{%
\tilde{L}+}1\right) /r^{2}$, more stronger (since, $0<\alpha ^{2}\leq 1$, $%
\xi \geq 0$, and $\tilde{\kappa}\geq 0$), whereas the WYMM strength, $\sigma 
$, makes it weaker (as documented in (\ref{e57}) along with (\ref{e58})).

We now follow the suggested conditionally exact solvability procedure discussed by Mustafa et al. \cite{ref701,Ref702} and use the substitution $\phi \left( r\right) =\exp \left( -\frac{\Omega }{2}r^{2}\right) \,R\left(r\right)$  to obtain
\begin{equation}
\left( r^{2}+\,\tilde{\kappa}\right) R^{\prime \prime }\left( r\right) +%
\left[ 2\left( 1-\Omega \,\tilde{\kappa}\right) r+\frac{\,\tilde{\kappa}}{r}%
-2\Omega r^{3}\right] R^{\prime }\left( r\right) +\left[ \mathcal{P}%
_{1}r^{2}-\mathcal{P}_{2}-\frac{2\xi \tilde{\kappa}}{r^{2}}\right] R\left(
r\right) =0,  \label{e60.1}
\end{equation}%
where%
\begin{equation}
\mathcal{P}_{1}=\tilde{E}^{2}+\Omega ^{2}\tilde{\kappa}-3\Omega ,\,\mathcal{P%
}_{2}=\mathcal{\tilde{L}}\left( \mathcal{\tilde{L}+}1\right) +2\Omega \,%
\tilde{\kappa}.  \label{e60.2}
\end{equation}%
We next use the change of variables $y=r^{2}$ to rewrite Eq.(\ref{e60.1}) as 
\begin{equation}
y\left( y+\,\tilde{\kappa}\right) R^{\prime \prime }\left( y\right) +\left[
\left( \frac{3}{2}-\Omega \,\tilde{\kappa}\right) y+\,\tilde{\kappa}-\Omega
y^{2}\right] R^{\prime }\left( y\right) +\left[ \mathcal{\tilde{P}}_{1}y-%
\mathcal{\tilde{P}}_{2}-\frac{\xi \tilde{\kappa}}{2y}\right] R\left(
y\right) =0,  \label{e61}
\end{equation}%
with%
\begin{equation}
\mathcal{\tilde{P}}_{i}=\frac{\mathcal{P}_{i}}{4}.  \label{e62}
\end{equation}%
One should notice that equation (\ref{e61}), as it stands, allows us to switch off EiBI-gravity (i.e., $\tilde{\kappa}=0$) and/or the Ricci scalar curvature manifested gravitational force field (i.e., $\xi =0$). Yet, it can be transformed into that of the confluent Heun differential equation (a procedure we feel very reluctant to recommend and/or use in the current methodical proposal) using a different change of variables, i.e., $z=-y/\tilde{\kappa}=-r^{2}/\tilde{\kappa}$, and therefore admits a general solution in the form of the confluent Heun functions%
\begin{eqnarray}
R(y) &=&A_{1}\,y^{\sqrt{\xi /2}}\,H_{C}(\Omega \tilde{\kappa},\sqrt{2\xi },-%
\frac{1}{2},-\frac{\tilde{\kappa}}{4}(3\Omega +4\mathcal{\tilde{P}}_{1}),%
\frac{\Omega \,\tilde{\kappa}}{2}-\mathcal{\tilde{P}}_{2}+\frac{\xi }{2}+%
\frac{1}{4},-\frac{y}{\tilde{\kappa}})  \notag \\
&&+A_{2}\,y^{-\sqrt{\xi /2}}\,H_{C}(\Omega \tilde{\kappa},-\sqrt{2\xi },-%
\frac{1}{2},-\frac{\tilde{\kappa}}{4}(3\Omega +4\mathcal{\tilde{P}}_{1}),%
\frac{\Omega \,\tilde{\kappa}}{2}-\mathcal{\tilde{P}}_{2}+\frac{\xi }{2}+%
\frac{1}{4},-\frac{y}{\tilde{\kappa}}).  \label{e63}
\end{eqnarray}%
Obviously, the finiteness of the wave function at $y=0$ (i.e., $r=0$) would suggest that $A_{2}=0$ and the wave function is, therefore, given by 
\begin{equation}
R(y)=A_{1}\,y^{\sqrt{\xi /2}}\,H_{C}(\Omega \tilde{\kappa},\sqrt{2\xi },-%
\frac{1}{2},-\frac{\tilde{\kappa}}{4}(3\Omega +4\mathcal{\tilde{P}}_{1}),%
\frac{\Omega \,\tilde{\kappa}}{2}-\mathcal{\tilde{P}}_{2}+\frac{\xi }{2}+%
\frac{1}{4},-\frac{y}{\tilde{\kappa}}).  \label{e64}
\end{equation}%
The truncation of the confluent Heun function $H_{C}\left( \alpha ,\beta
,\gamma ,\delta ,\eta ,z\right) $ into a polynomial of order $n_{r}+1;\,n_{r}\geq 0$, is secured by the condition (c.f., e.g., \cite%
{ref701,ref71}) that%
\begin{equation}
\delta =-\alpha \left( n_{r}+\frac{1}{2}\left[ \beta +\gamma +2\right]
\right) \Longrightarrow \mathcal{\tilde{P}}_{1}=\Omega \left( n_{r}+\sqrt{%
\frac{\xi }{2}}\right) .  \label{e64.1}
\end{equation}%
This would in turn yield%
\begin{equation}
\tilde{E}^{2}=2\Omega \left( 2n_{r}+\sqrt{2\xi }+\frac{3}{2}\right) -\Omega
^{2}\,\tilde{\kappa}\Longleftrightarrow E=\pm \sqrt{m_{\circ }^{2}+2\Omega
\alpha ^{2}\left( 2n_{r}+\sqrt{2\xi }+3\right) }.  \label{e65}
\end{equation}%
In what follows, however, we shall use our change of variables, $y=r^{2}$, along with (\ref{e61}), to show that the result in (\ref{e64.1}), hence (\ref%
{e65}), is based on some conditions/correlations that facilitates the truncation procedure above and render the solution to be classified as a \textit{conditionally exact solution}, therefore. Some of these conditions/correlations were discussed by Ishkhanyan et al. \cite{ref71}. We shall add one more condition/correlation in the sequel. Consequently, this result (\ref{e65}) has to be rechecked and elaborated through the following power series expansion procedure.

\section{Parametric correlation/condition associated with such a solution}

We shall now identify one feasible/admissible  correlation/condition that should be taken into
account while using the result in (\ref{e65}), for the truncation condition of the confluent Heun $\left( n_{r}+1\right) $-polynomial (\ref{e64.1}). In so doing, we closely follow our recipe in \cite{ref701} and use%
\begin{equation}
R\left( y\right) =y^{\nu }\sum\limits_{j=0}^{\infty }C_{j}\,y^{j},
\label{e66}
\end{equation}%
in (\ref{e61}) to obtain, in a straightforward manner, the relations 
\begin{equation}
C_{0}\left[ \tilde{\kappa}\nu ^{2}-\frac{\xi \tilde{\kappa}}{2}\right]
=0,\;C_{0}\left[ \nu \left( \nu +\frac{1}{2}-\Omega \,\tilde{\kappa}\right) -%
\mathcal{\tilde{P}}_{2}\right] +C_{1}\left[ \tilde{\kappa}\left( \nu
+1\right) ^{2}-\frac{\xi \tilde{\kappa}}{2}\right] =0,  \label{e68}
\end{equation}%
and%
\begin{equation}
C_{j+2}\left[ \tilde{\kappa}\left( j+\nu +2\right) ^{2}-\frac{\xi \tilde{%
\kappa}}{2}\right] =C_{j}\,\left[ \Omega \left( j+\nu \right) -\mathcal{%
\tilde{P}}_{1}\right] +C_{j+1}\left[ \mathcal{\tilde{P}}_{2}-\left( j+\nu
+1\right) \left( j+\nu +\frac{3}{2}-\Omega \,\tilde{\kappa}\right) \right] .
\label{e69}
\end{equation}%
One should notice, at this point, that our change of variable $y=r^{2}$ allows us to consider some special cases, to be discussed in the sequel, like: (i) KG-oscillators in EiBI-gravity at $\tilde{\kappa}\neq 0$ with the Ricci scalar curvature effect at $\xi \neq 0$ (which is the core of the current proposal),  and (ii) KG-oscillators without EiBI-gravity, $\tilde{\kappa}=0$, but with the scalar curvature effect, $\xi \neq0$ . These are feasibly viable and interesting models for quantum gravity and astroparticle physics.

\subsection{KG-oscillators in EiBI-gravity, $\tilde{\kappa}\neq 0$, with the Ricci scalar curvature effect }

Notably, for the Eddington parameter $\tilde{\kappa}\neq 0$ and the scalar curvature parameter $\xi \neq 0$, the first condition in 
(\ref{e68}) would suggest that since $C_{0}\neq 0$ we have 
\begin{equation}
\left[ \tilde{\kappa}\nu ^{2}-\frac{\xi \tilde{\kappa}}{2}\right]
=0\Longrightarrow \nu =\sqrt{\frac{\xi }{2}}.  \label{70}
\end{equation}%
The second condition in (\ref{e68}), on the other hand, would imply that%
\begin{equation}
C_{1}=\frac{\mathcal{\tilde{P}}_{2}-\nu \left( \nu +\frac{1}{2}-\Omega \,%
\tilde{\kappa}\right) }{\tilde{\kappa}\left( 2\nu +1\right) }C_{0};\;C_{0}=1.
\label{e71}
\end{equation}%
Next, we truncate our power series in (\ref{e66}) into a polynomial of order  $n_{r}+1=n\geq 1$ so that $\forall j=n_{r}$ we require $C_{n_{r}+2}=0$, $%
C_{n_{r}+1}\neq 0$, and $C_{n_{r}}\neq 0$. However, to facilitate the so called \textit{conditionally exact solvability}, of the problem at hand, one would enforce the valid/viable assumptions that, since 
$C_{n_{r}+1}\neq 0$ and $C_{n_{r}}\neq 0$ we may very well assume $\mathcal{\tilde{P}}_{2}-\left(
n_{r}+\nu +1\right) \left( n_{r}+\nu +\frac{3}{2}-\Omega \,\tilde{\kappa}%
\right) =0$ and $\Omega \left( n_{r}+\nu \right) -\mathcal{\tilde{P}}_{1}=0$. Under such assumptions, one would, respectively, obtain%
\begin{equation}
\Omega \,\tilde{\kappa}=\frac{\alpha ^{2}\left[ \left( 2n_{r}+\sqrt{2\xi }%
\right) \left( 2n_{r}+\sqrt{2\xi }+5\right) +6\right] +\sigma ^{2}-\left[
\ell \left( \ell +1\right) +2\xi \left( 1-\alpha ^{2}\right) \right] }{%
2\alpha ^{2}\left( 2n_{r}+\sqrt{2\xi }+4\right) },  \label{e72}
\end{equation}%
and%
\begin{equation}
\tilde{E}^{2}=2\Omega \left( 2n_{r}+\sqrt{2\xi }+\frac{3}{2}\right) -\Omega
^{2}\,\tilde{\kappa}\Longleftrightarrow E=\pm \sqrt{m_{\circ }^{2}+2\Omega
\alpha ^{2}\left( 2n_{r}+\sqrt{2\xi }+3\right) }.  \label{e73}
\end{equation}%
This result is in exact accord with that in (\ref{e65}) obtained from the truncation condition (\ref{e64.1}) of the confluent Heun functions in (\ref%
{e64}). However, equation (\ref{e72}) identifies a parametric correlation
between the oscillator frequency $\Omega $ and the Eddington parameter $%
\tilde{\kappa}$. Which, in effect, mandates that $\Omega \,\neq 0\neq \tilde{%
\kappa}$ (i.e., neither $\tilde{\kappa}$ nor $\Omega $ are allowed to take zero values). Moreover, the restriction, by definition, $\Omega \,\tilde{\kappa}>0$ would result that our orbital angular momentum quantum number $0\leq \ell =0,1,2,\cdots $ is constrained by the relation 
\begin{equation}
\ell \left( \ell +1\right) <\alpha ^{2}\left[ \left( 2n_{r}+2\nu \right)
\left( 2n_{r}+2\nu +5\right) +6\right] +\sigma ^{2}-2\xi \left( 1-\alpha
^{2}\right) ,  \label{e74}
\end{equation}%
to imply that%
\begin{equation}
0\leq \ell <-\frac{1}{2}+\sqrt{\frac{1}{4}+\alpha ^{2}\left[ \left( 2n_{r}+\sqrt{%
2\xi }\right) \left( 2n_{r}+\sqrt{2\xi }+5\right) +6\right] +\sigma
^{2}-2\xi \left( 1-\alpha ^{2}\right) }.  \label{e75}
\end{equation}%
are the only allowed $\ell $-states. This is a price one has, some times, to pay for \textit{conditionally exact solvabilities} of some complicated systems like the one at hand. In light of the above power series experience, we, moreover, observe that the truncation condition (\ref{e64.1}) on the confluent Heun function $H_{C}\left( \alpha ,\beta ,\gamma ,\delta
,\eta ,z\right) $ into a polynomial of order $n_{r}+1$ should be associated with the correlation in (\ref{e72}). This is not only interesting, but also mandatory while using the confluent Heun functions. Consequently, our radial wavefunction would read%
\begin{equation}
R\left( y\right) =y^{\sqrt{\xi /2}}\sum\limits_{j=0}^{n_{r}+1}C_{j}y^{j}%
\Leftrightarrow R\left( r\right) =r^{\sqrt{2\xi }}\sum%
\limits_{j=0}^{n_{r}+1}C_{j}\,r^{2j},  \label{e76}
\end{equation}%
where $C_{0}$, $C_{1}$ are given in (\ref{e71}), and 
\begin{eqnarray}
    C_{j+2}\left[ \tilde{\kappa}\left( j+\sqrt{\frac{\xi }{2}}+2\right) ^{2}-%
\frac{\xi \tilde{\kappa}}{2}\right] &=& C_{j}\,\left[ \Omega \left( j+\sqrt{%
\frac{\xi }{2}}\right) -\mathcal{\tilde{P}}_{1}\right] \notag \\
&+&C_{j+1}\left[ 
\mathcal{\tilde{P}}_{2}-\left( j+\sqrt{\frac{\xi }{2}}+1\right) \left( j+%
\sqrt{\frac{\xi }{2}}+\frac{3}{2}-\Omega \,\tilde{\kappa}\right) \right].
\label{e77}
\end{eqnarray}
The duty of which is to allow one to obtain $C_2,\,C_3,\,\cdots$. For example, for $j=0$, one obtains%
\begin{equation}
C_{2}\left[ \tilde{\kappa}\left( \sqrt{\frac{\xi }{2}}+2\right) ^{2}-\frac{%
\xi \tilde{\kappa}}{2}\right] =C_{0}\,\left[ \Omega \left( \sqrt{\frac{\xi }{%
2}}\right) -\mathcal{\tilde{P}}_{1}\right] +C_{1}\left[ \mathcal{\tilde{P}}%
_{2}-\left( \sqrt{\frac{\xi }{2}}+1\right) \left( \sqrt{\frac{\xi }{2}}+%
\frac{3}{2}-\Omega \,\tilde{\kappa}\right) \right] ,  \label{e78}
\end{equation}%
and for $j=1$, we get
\begin{equation}
C_{3}\left[ \tilde{\kappa}\left( \sqrt{\frac{\xi }{2}}+3\right) ^{2}-\frac{%
\xi \tilde{\kappa}}{2}\right] =C_{1}\,\left[ \Omega \left( 1+\sqrt{\frac{\xi 
}{2}}\right) -\mathcal{\tilde{P}}_{1}\right] +C_{2}\left[ \mathcal{\tilde{P}}%
_{2}-\left( \sqrt{\frac{\xi }{2}}+2\right) \left( \sqrt{\frac{\xi }{2}}+%
\frac{5}{2}-\Omega \,\tilde{\kappa}\right) \right] ,  \label{e79}
\end{equation}%
and so on. Notice that the term $y^{\sqrt{\xi /2}}$ in (\ref{e76}) resembles the asymptotic convergence of the radial wave function in (\ref%
{e64}) as $y\rightarrow 0$ (i.e., $r\rightarrow 0$).
\begin{figure}[ht!]  
\centering
\includegraphics[width=0.35\textwidth]{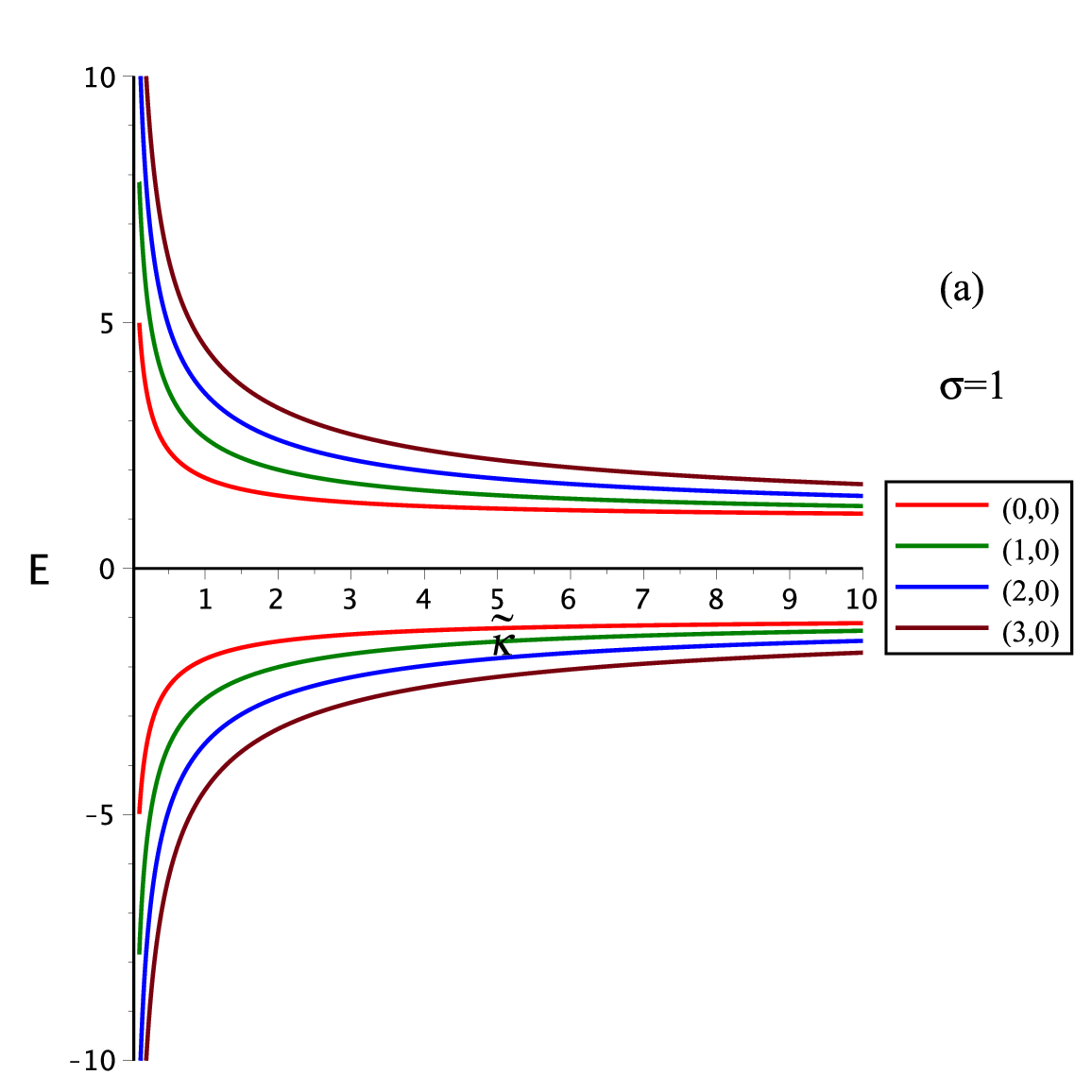}
\includegraphics[width=0.35\textwidth]{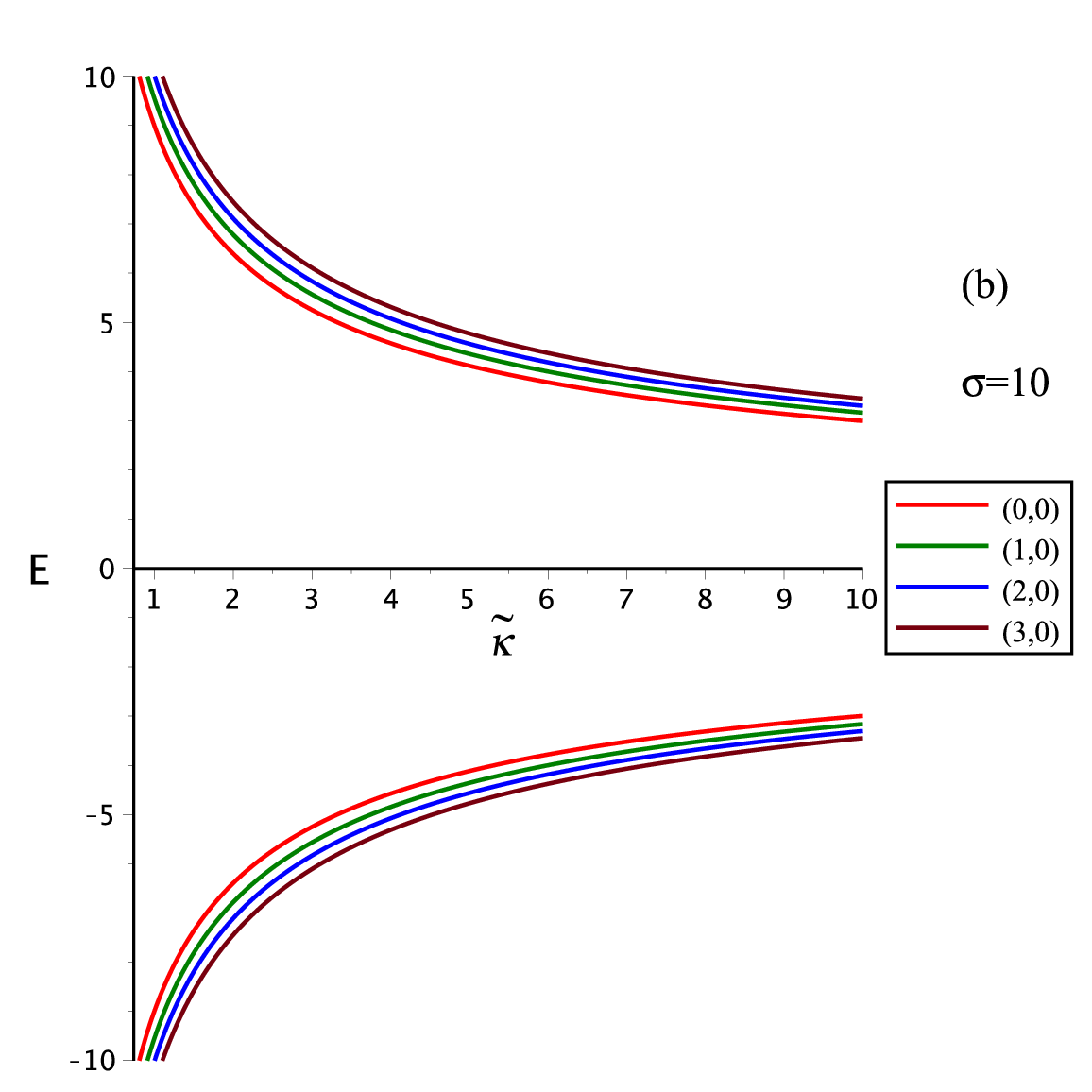}
\includegraphics[width=0.35\textwidth]{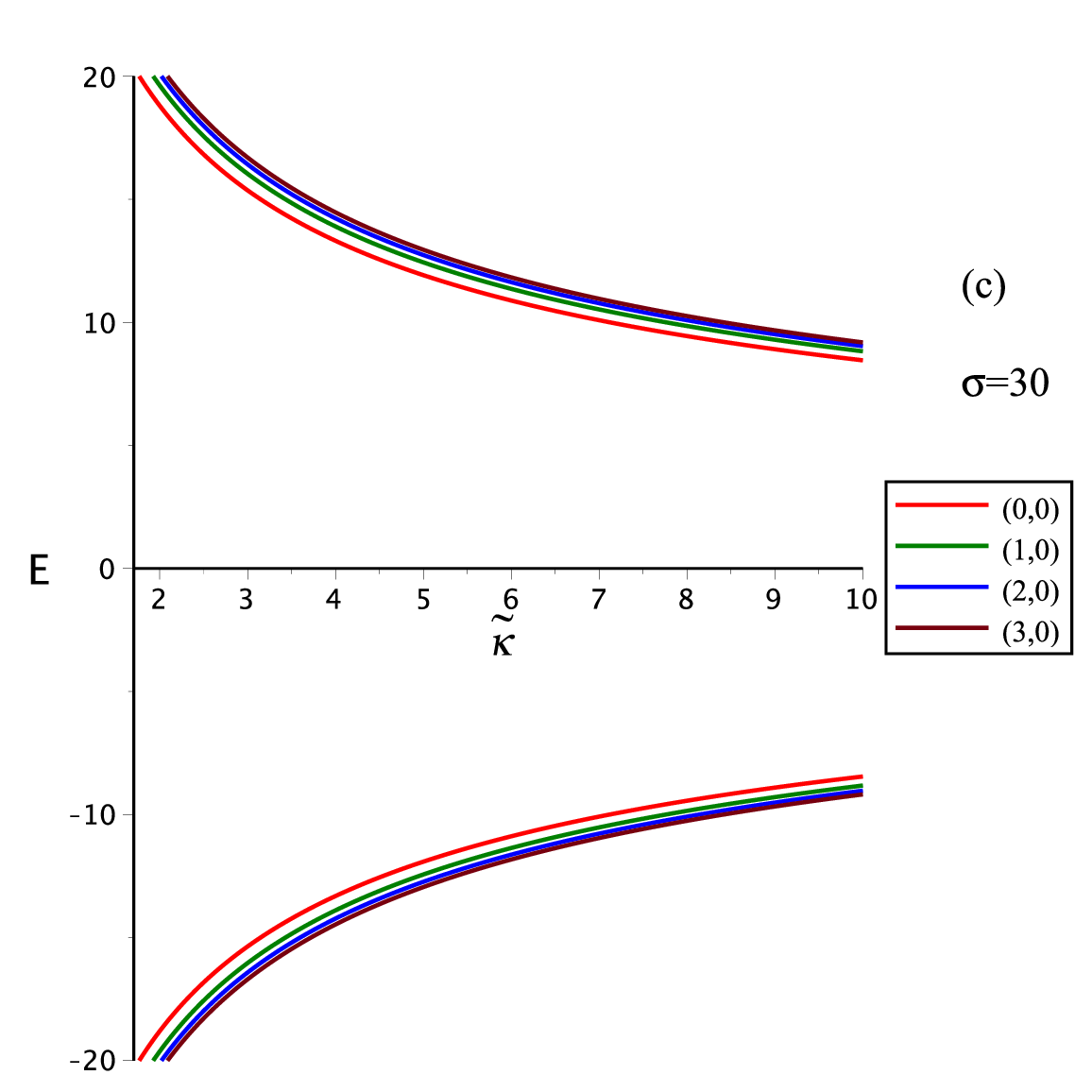}
\includegraphics[width=0.35\textwidth]{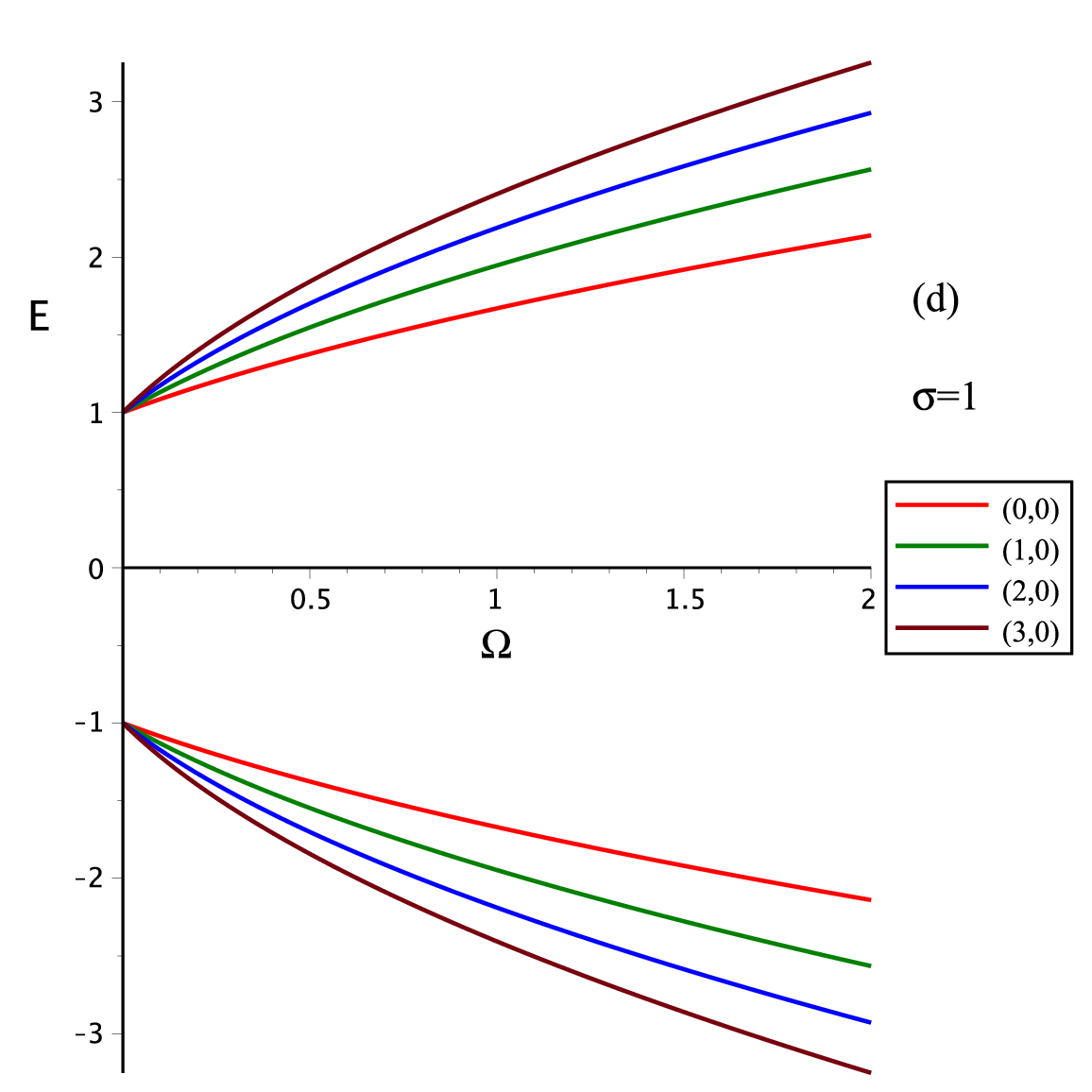}
\caption{\small 
{ The energy levels in Eq.s (\ref{e73}), along with (\ref{e72}%
), for the KG-oscillators in EiBI-gravity with the Ricci scalar curvature and a WYMM. The corresponding $\left( n_{r},\ell \right) $- states, for $%
\ell =0$ and $n_{r}=0,1,2,3$, are plotted for $m_{\circ }=1$, $\alpha =0.5$, $\xi =1/6$
(a) at $\sigma =1$ against different Eddington parameter $\tilde{%
\kappa}>0$ values, (b) at $\sigma =10$ against different Eddington parameter $\tilde{%
\kappa}>0$ values, (c) at $\sigma =30$ against different Eddington parameter $\tilde{%
\kappa}>0$ values, and (d) at $\sigma =1$, against different KG-oscillators' frequencies $\Omega >0$.}}
\label{fig1}
\end{figure}%
\begin{figure}[ht!]  
\centering
\includegraphics[width=0.35\textwidth]{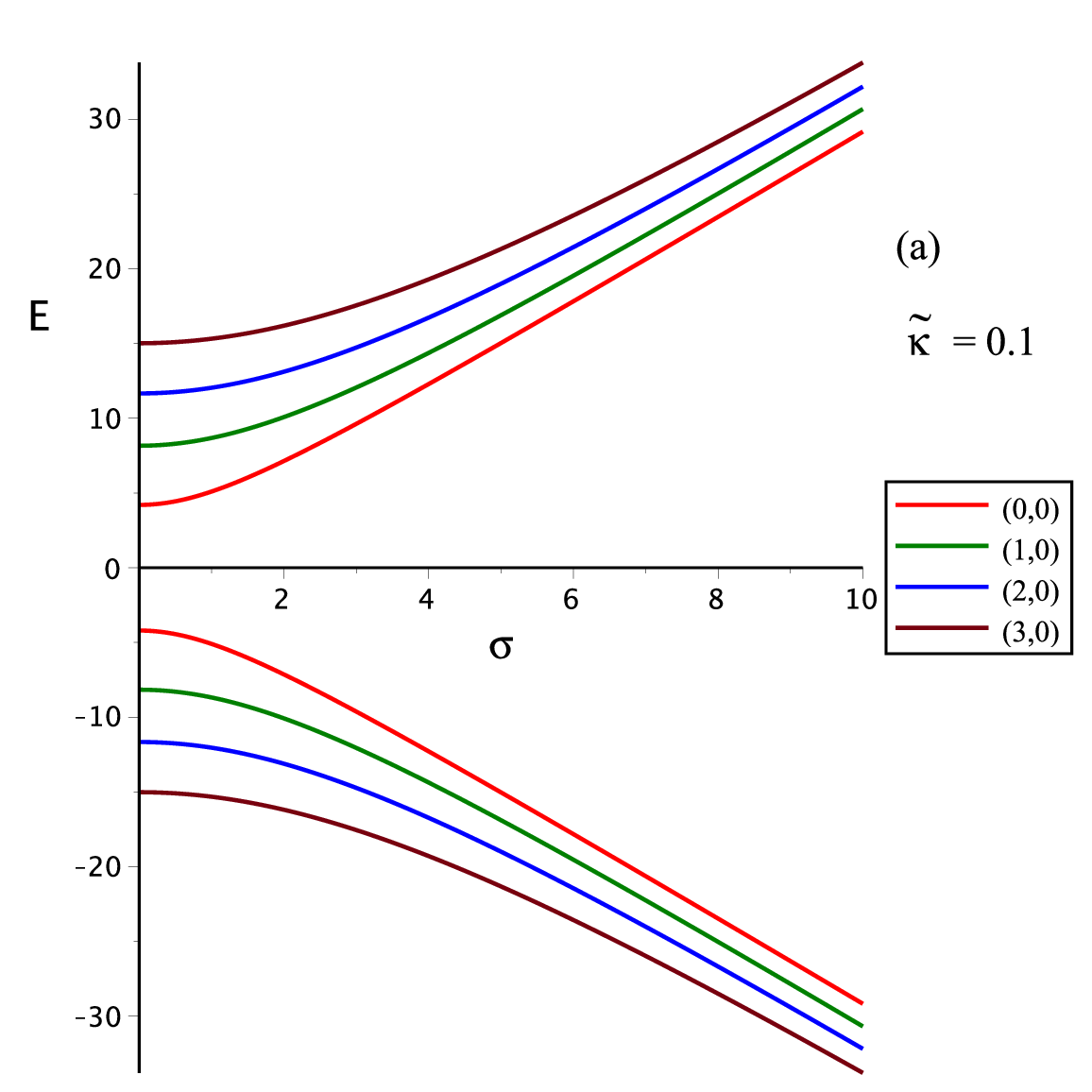}
\includegraphics[width=0.35\textwidth]{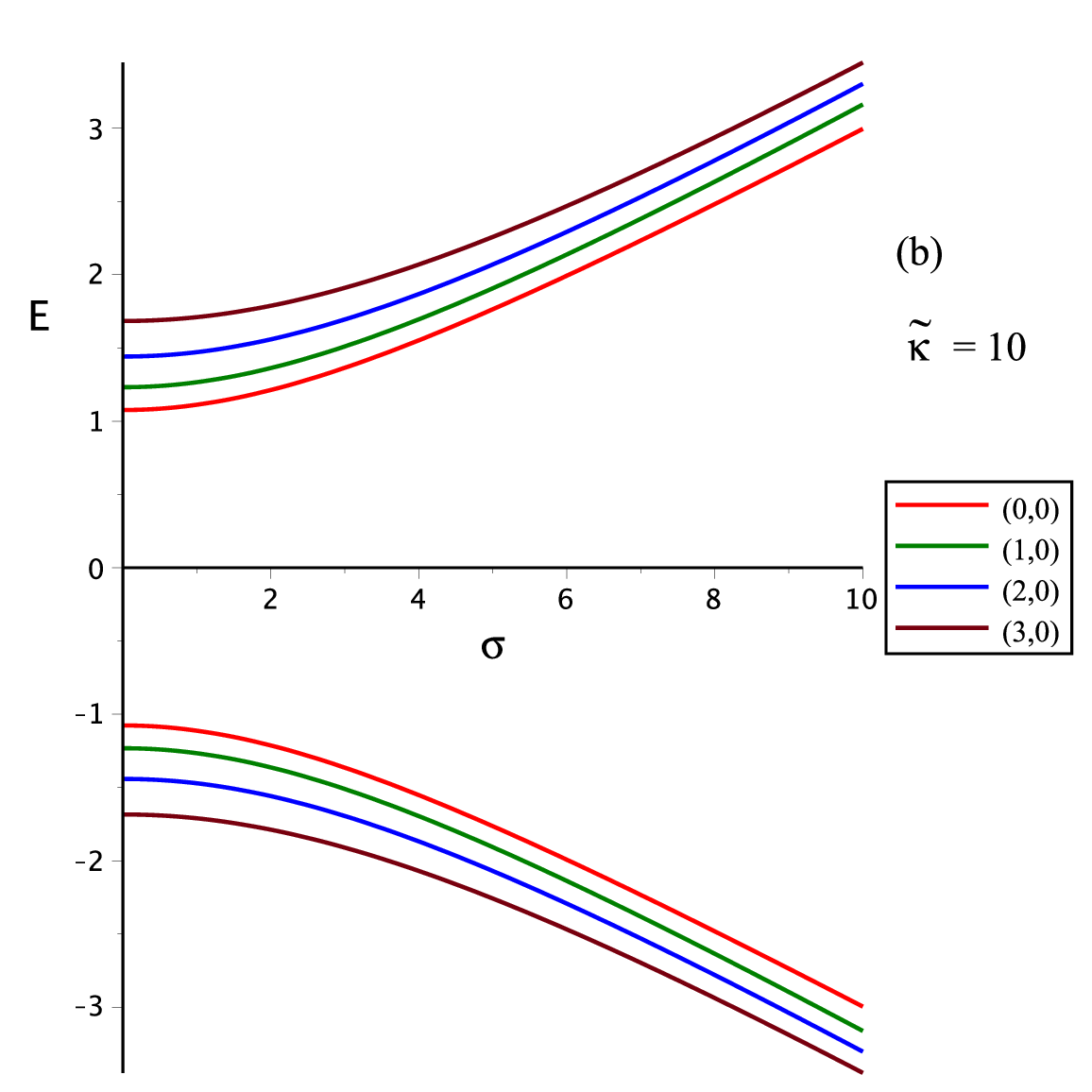}
\caption{\small 
{ The energy levels in Eq.s (\ref{e73}), along with (\ref{e72}%
), for the KG-oscillators in EiBI-gravity with the Ricci scalar curvature and a WYMM. The corresponding $\left( n_{r},\ell \right) $- states, for
different $\ell $ and $n_{r}$ values, for $\ell =0$ and $n_{r}=0,1,2,3$, at $m_{\circ }=1$, $\alpha
=0.5$, $\xi =1/6$ (a) at $\tilde{\kappa}=0.1$  against WYMM strength $\sigma \geq 0 $ values, and (b) at $\tilde{\kappa}=10$ against
against WYMM strength $\sigma \geq 0 $ values.}}
\label{fig2}
\end{figure}%

In Figure 1, we plot the energy levels in Eq.s (\ref{e73}), along with (\ref{e72}), for the KG-oscillators in EiBI-gravity with the Ricci scalar curvature and a WYMM. The corresponding $\left( n_{r},\ell \right) $- states, for $%
\ell =0$ and $n_{r}=0,1,2,3$,  for $m_{\circ }=1$, $\alpha =0.5$, and $\xi =1/6$.  The energy levels against different Eddington parameter $\tilde{%
\kappa}>0$ values are shown for  $\sigma =1$,  $\sigma =10$, and $\sigma =30$, in 1(a), 1(b), and 1(c), respectively. In 1(d) we show the energy levels at $\sigma =1$, against different frequencies of KG-oscillators $\Omega >0$. In Figure 2,  the energy levels are plotted against the WYMM  strength $\sigma \geq 0$ at $\tilde{\kappa}=0.1$ and $\tilde{\kappa}=10$ in 2(a) and 2(b), respectively. Whereas, for different $\left( n_{r},\ell \right) $- states, the energies are plotted against different Eddington parameter $\tilde{%
\kappa}>0$ values at WYMM strengths $\sigma =1$ and $\sigma =10$ (in 3(a) and 3(b), respectively), and  against different WYMM strength $\sigma\geq 0$ at the Eddington parameter $\tilde{\kappa}=0.1$ and  $\tilde{\kappa}=10$, (3(c) and 3(d), respectively).

Obviously, all the figures indicate that the energy gap between positive and negative energy levels is preserved. This energy gap is observed to decrease with increasing Eddington parameter $\tilde{\kappa}$ (as in Figs.
1(a), 1(b), 1(c), 3(a) and 3(b)), where clustering of the energy levels, around $E=\pm m_{\circ }
$, is observed eminent for extreme Eddington gravity, $\tilde{\kappa}>>1$. This is, in fact, the asymptotic tendency of the energy levels in (\ref{e73}), along with (\ref%
{e72}), at $\tilde{\kappa}>>1$. The energy gap, on the other hand, increases with increasing KG-oscillators' frequency $\Omega $ (as in Fig. 1(d)), and with increasing WYMM strength $\sigma $ (as in Figs. 2(a), 2(b), 3(c) and 3(d)). Moreover, one observes that the separation between the energy levels as well as the KG-oscillator energies decrease with increasing values of the WYMM (documented in 1(a), 1(b), 1(c), 1(d), 3(a), and 3(b)). One would anticipate that at extreme WYMM strength (i.e., $\sigma>>1$) all quantum states would gather and collapse into one state (documented in 1(a), 1(b), 1(c), 3(a) and 3(b)). 

At this point, one should be aware that the truncation of the confluent Heun function into a polynomial of order $n_{r}+1$ should be associated with some correlations/conditions. The use of such correlations is mandatory. We have just presented one of the many feasible correlations, to come out with a \textit{conditionally exact solution}, in addition to those reported by Ishkhanyan \cite{ref71}. In what follows we shall show how quite flexible and handy the usage of the above methodical proposal is. We do so through the following
illustrative examples.
\begin{figure}[ht!]  
\centering
\includegraphics[width=0.35\textwidth]{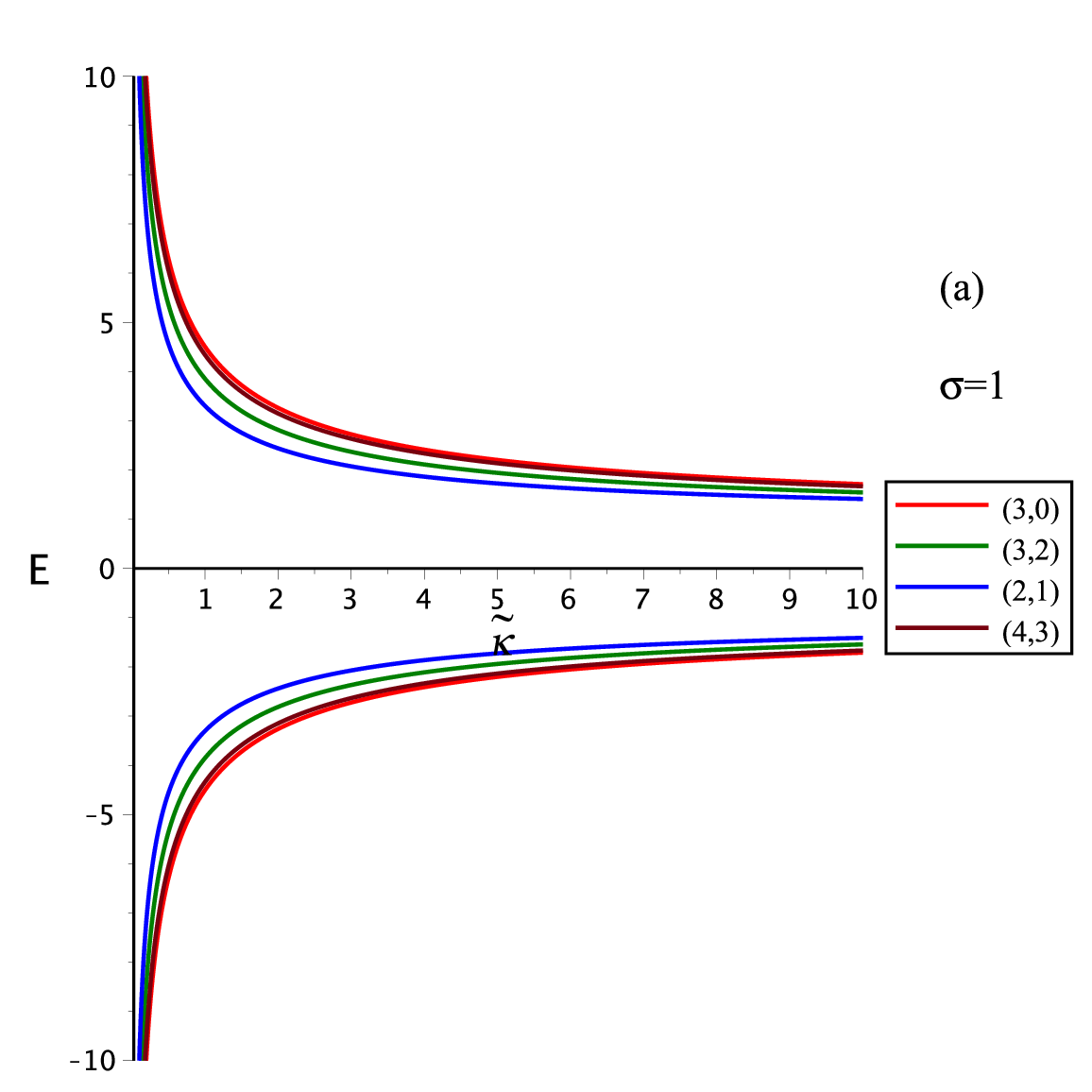}
\includegraphics[width=0.35\textwidth]{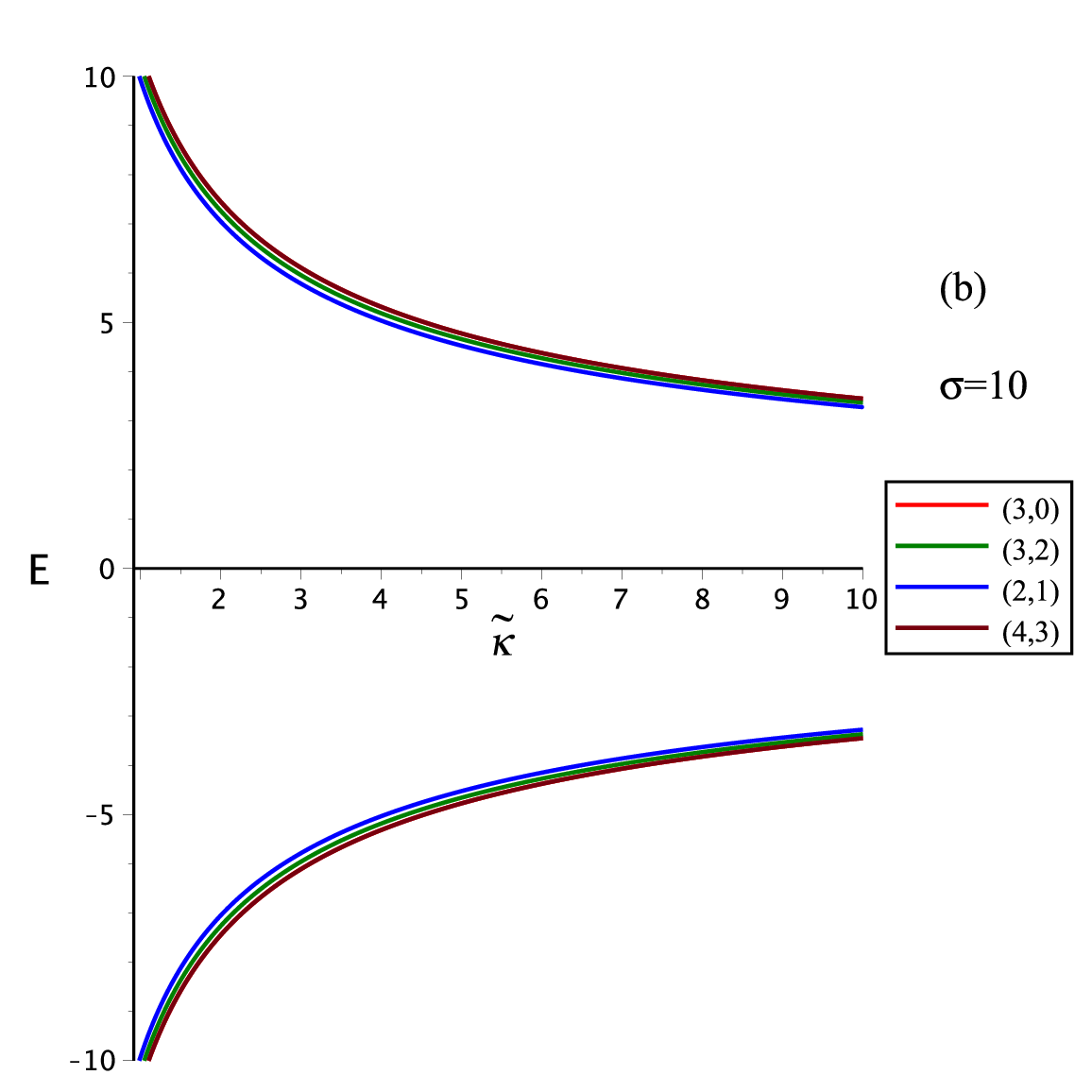}
\includegraphics[width=0.35\textwidth]{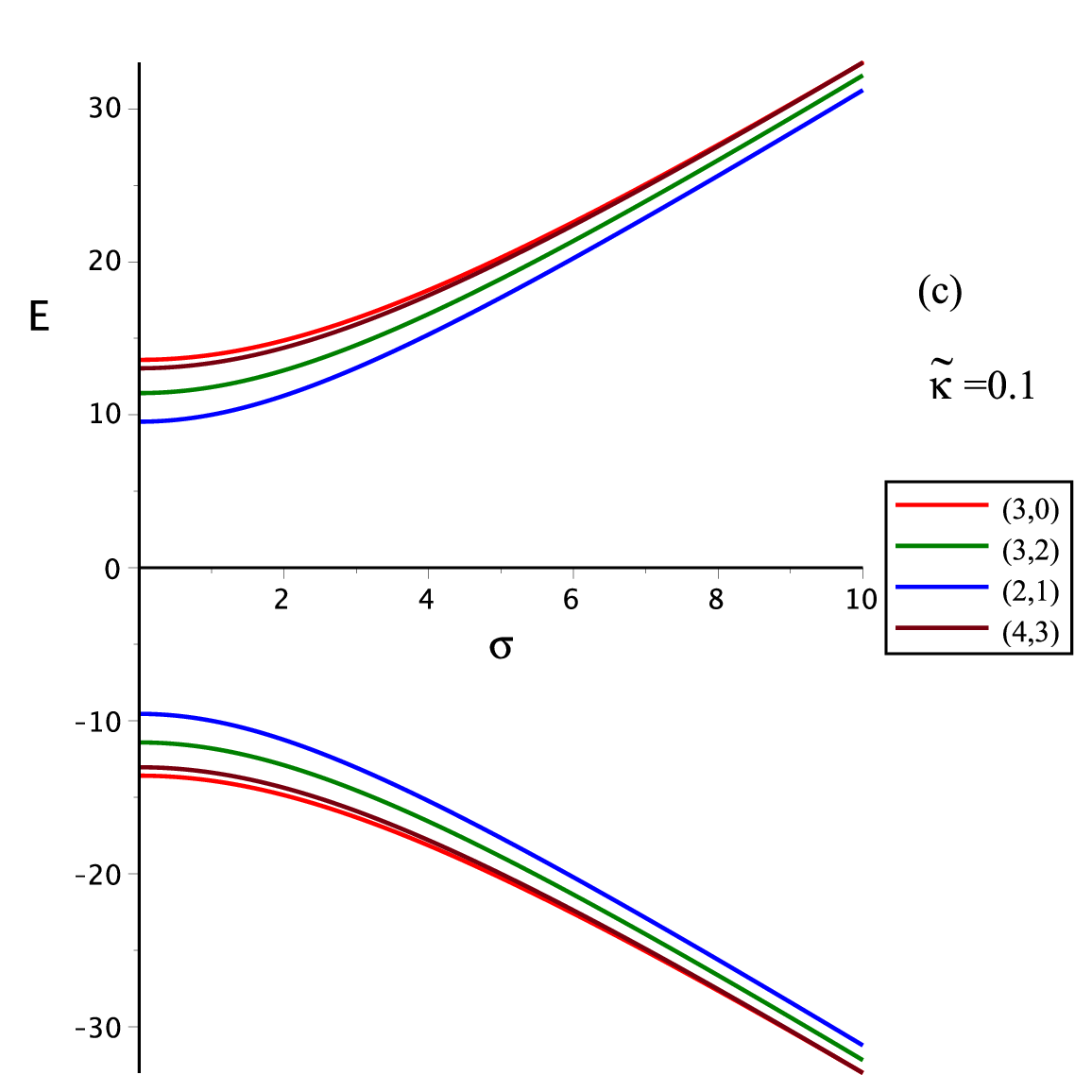}
\includegraphics[width=0.35\textwidth]{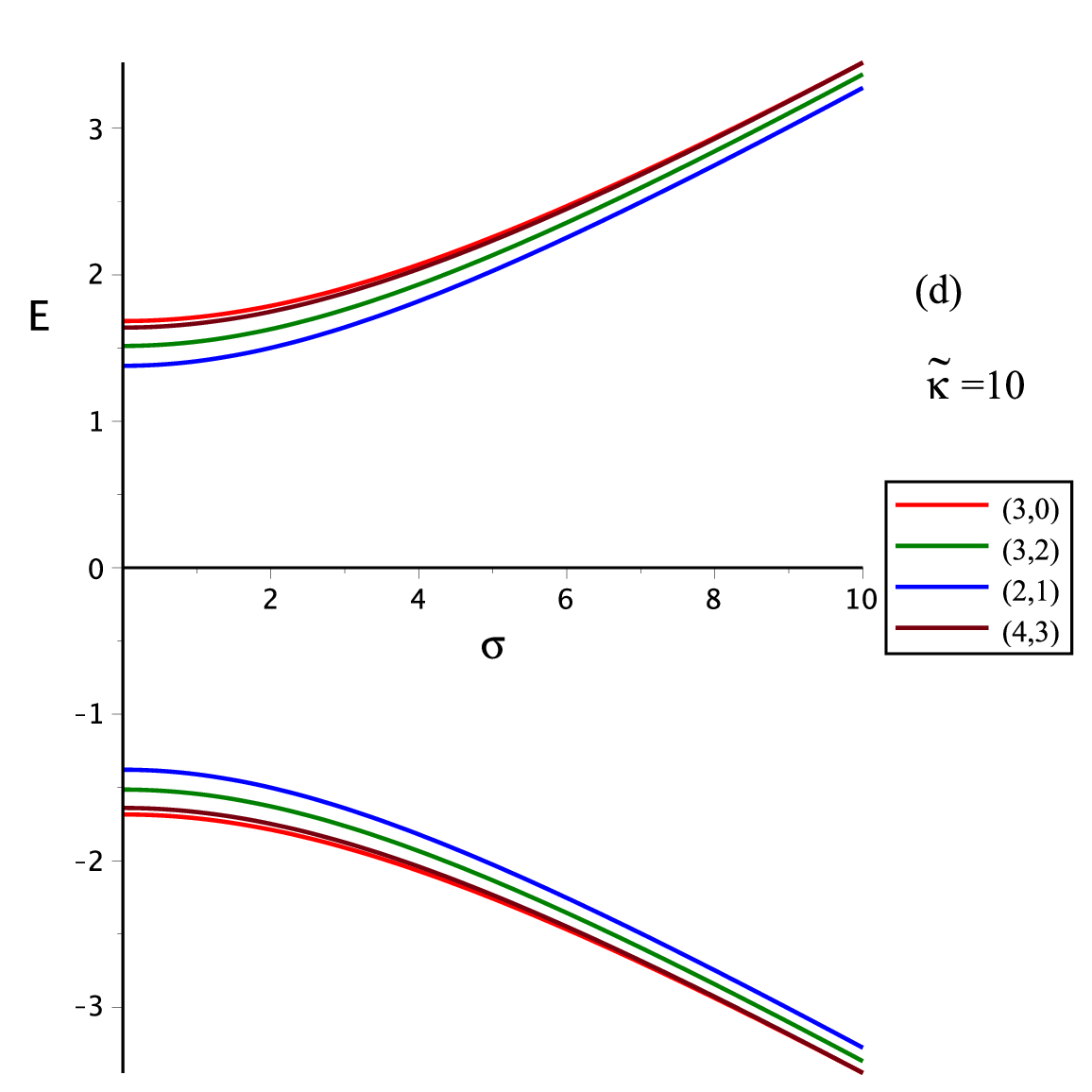}
\caption{\small 
{ The energy levels in Eq.s (\ref{e73}), along with (\ref{e72}%
), for the KG-oscillators in EiBI-gravity with the Ricci scalar curvature and a WYMM. The energies are for different $\left( n_{r},\ell \right) $- states, for $m_{\circ }=1$, $\alpha =0.5$, $\xi =1/6$
(a) at $\sigma =1$,  against different Eddington parameter $\tilde{%
\kappa}>0$ values, (b) at $\sigma =10$ against different Eddington parameter $\tilde{%
\kappa}>0$ values, (c) at $\tilde{\kappa}=0.1$ against different WYMM strength $\sigma\geq 0$ values, and (d) at $\tilde{\kappa}=10$ against different WYMM strength $\sigma \geq 0$ values.}}
\label{fig3}
\end{figure}%
Moreover, the L.H.S. of (\ref{e69}) is zero to yield%
\begin{equation}
C_{j+1}\left[ \mathcal{\tilde{P}}_{2}-\left( j+\nu +1\right) \left( j+\nu +%
\frac{3}{2}-\Omega \,\tilde{\kappa}\right) \right] =C_{j}\,\left[ \mathcal{%
\tilde{P}}_{1}-\Omega \left( j+\frac{\mathcal{\tilde{L}}}{2}\right) \right] 
\label{c4}
\end{equation}%
We may now truncate the power series into a polynomial of order $n_{r}\geq 0$ by requiring that $\forall j=n_{r}$ we have $C_{n_{r}+1}=0$, and $%
C_{n_{r}}\neq 0$. Consequently,%
\begin{equation}
\mathcal{\tilde{P}}_{1}=\Omega \left( n_{r}+\frac{\mathcal{\tilde{L}}}{2}%
\right) \Longrightarrow \tilde{E}^{2}=2\Omega \left( 2n_{r}+\mathcal{\tilde{L%
}+}\frac{3}{2}\right) \Longrightarrow E=\pm \sqrt{2\Omega \alpha ^{2}\left(
2n_{r}+\mathcal{\tilde{L}+}3\right) +m_{\circ }^{2}}  \label{c5}
\end{equation}%
This special case represents KG-oscillators in GM-spacetime and a WYMM and is in exact accord that in Eq.(31) of \cite{ref017}. Moreover, the radial wave function reads%
\begin{equation}
R\left( y\right) =y^{\mathcal{\tilde{L}}/2}\sum%
\limits_{j=0}^{n_{r}}C_{j}y^{j}\Leftrightarrow R\left( r\right) =r^{%
\mathcal{\tilde{L}}}\sum\limits_{j=0}^{n_{r}}C_{j}\,r^{2j}  \label{c05.1}
\end{equation}

\subsection{KG-oscillators in GM-spacetime without EiBI-gravity, $\tilde{%
\kappa}=0$, and with the Ricci scalar curvature effect}

Without EiBI-gravity, $\tilde{\kappa}=0$, in the presence of Ricci scalar curvature, $\xi \neq 0$, we use (\ref{e66}) in equation (\ref{e61}) to obtain, with $C_{0}\neq 0$,%
\begin{equation}
\nu \left( \nu +\frac{1}{2}\right) -\mathcal{\tilde{P}}_{2}=0\Longrightarrow
\nu =\frac{\mathcal{\tilde{L}}}{2},  \label{c5.2}
\end{equation}%
where%
\begin{equation}
\mathcal{\tilde{L}}=-\frac{1}{2}+\sqrt{\frac{1}{4}+\frac{\ell \left( \ell
+1\right) -\sigma ^{2}+2\xi \left( 1-\alpha ^{2}\right) }{\alpha ^{2}}}.
\label{c5.3}
\end{equation}%
Consequently, the truncation of our power series into a polynomial of order $%
n_{r}$ is secured by the requirement that $\forall j=n_{r}$ we have $%
C_{n_{r}+1}=0$, $C_{n_{r}}\neq 0$, and therefor4e%
\begin{equation}
\mathcal{\tilde{P}}_{1}-\Omega \left( n_{r}+\nu \right) =0\Longrightarrow 
\frac{\tilde{E}^{2}-3\Omega }{4}=\Omega \left( n_{r}+\frac{\mathcal{\tilde{L}%
}}{2}\right) \Longrightarrow \frac{E^{2}-m_{\circ }^{2}}{\alpha ^{2}}%
-6\Omega =4\Omega \left( n_{r}+\frac{\mathcal{\tilde{L}}}{2}\right) ,
\label{c5.4}
\end{equation}%
to imply%
\begin{equation}
E=\pm \sqrt{m_{\circ }^{2}+4\Omega \alpha ^{2}\left( n_{r}+\frac{\sqrt{%
\alpha ^{2}+4\ell \left( \ell +1\right) -4\sigma ^{2}+8\xi \left( 1-\alpha
^{2}\right) }}{4\alpha }+\frac{5}{4}\right) }.  \label{c5.5}
\end{equation}%
This result is in exact agreement with that reported in Eq.(38) by Bragan\c{c}a et al. \cite{Re11} without the WYMM (i.e., for $\sigma =0$). Again, the radial wave function reads%
\begin{equation}
R\left( y\right) =y^{\mathcal{\tilde{L}}/2}\sum%
\limits_{j=0}^{n_{r}}C_{j}y^{j}\Leftrightarrow R\left( r\right) =r^{%
\mathcal{\tilde{L}}}\sum\limits_{j=0}^{n_{r}}C_{j}\,r^{2j}  \label{c5.6}
\end{equation}

\section{Concluding remarks}

We have studied and investigated KG-oscillators in a GM spacetime in EiBI-gravity, including a WYMM and the Ricci scalar $R=R_{\upsilon }^{\upsilon }$ effects. In light of our investigation, our observations are in order.

In connection with the quantum mechanical central repulsive core, $\mathcal{\ell }\left( \ell
+1\right) /r^{2}\longrightarrow \mathcal{\tilde{L}}\left( \mathcal{\tilde{L}+%
}1\right) /r^{2}$,  where $0\leq\mathcal{\tilde{L}}\in \mathbb{R}$ identifies a new irrational orbital quantum number, we recollect (from Eq.  (\ref{e58})) that %
\begin{equation}
\frac{\mathcal{\tilde{L}}\left( \mathcal{\tilde{L}+}1\right)}{r^2} =%
\frac{\ell \left( \ell +1\right) -\sigma ^{2}+2\xi \left( 1-\alpha
^{2}\right) }{\alpha ^{2} r^2}+\frac{2\Omega \,\tilde{\kappa}}{r^2}. \label{c13}
\end{equation}%
It is clear that the inclusion of the Ricci scalar curvature $R$ and EiBI-gravity,   manifestly and effectively, introduces additional repulsive force fields, $\frac{2\xi \left( 1-\alpha
^{2}\right) }{\alpha ^{2} r^2}$ and $\frac{2\Omega \,\tilde{\kappa}}{r^2}$,  respectively. However, an attractive force field is introduced by the WYMM,  $-\frac{\sigma^2}{r^2}$. The competitions between such fields identifies the allowed orbital $\ell$ excitations, provided that $0\leq\ell \in \mathbb{R}$, as documented in (\ref{e75}).  In addition, one should keep in mind that $\mathcal{\tilde{L}}=\ell$ for $\sigma=\xi=\tilde{\kappa}=0$ and $\alpha=1$ (i.e., flat Minkowski spacetime). 

As to the effects on the spectroscopic structure of the KG-oscillators in a GM spacetime in EiBI-gravity, in a WYMM and the Ricci scalar curvature, we have observed that the energy gap, between positive and negative energy levels, decreases with increasing Eddington parameter $\tilde{\kappa}$ (documented in Fig.s 1(a), 1(b), 1(c), 3(a) and 3(b)). Yet, the energy levels tend to cluster around $E=\pm m_{\circ }$ for extreme Eddington gravity (i.e.,  $\tilde{\kappa}>>1$), which, in fact, reflects the asymptotic tendency of the energy levels in (\ref{e73}), along with (\ref%
{e72}), at $\tilde{\kappa}>>1$. 
Whereas, the energy gap increases with increasing KG-oscillators' frequency $\Omega $ (documented in Fig. 1(d)), as well as with increasing WYMM strength $\sigma $ ( documented in Fig.s 2(a), 2(b), 3(c), and 3(d)). Moreover, we have observed that the separation between the energy levels (among positive or among negative energy states) as well as the KG-oscillator energies decrease  with increasing values of the WYMM strength $\sigma$ (documented in 1(a), 1(b), 1(c), 1(d), 3(a), and 3(b)). One would anticipate that at extreme WYMM strength (i.e., $\sigma>>1$) all quantum states would cluster and collapse into one state (documented in 1(a), 1(b), 1(c),  3(a), and 3(b)).  Yet, one may  observe the competition between  EiBI-gravitational field and the gravitational field introduced by  Ricci scalar curvature in Fig.s 2(a) and 2(b). Whilst the increase in the Eddington parameter  (i.e., stronger Eddington gravity) $\tilde{\kappa}>>1$ lowers the particle/anti-particle energies, the increase in the  WYMM strength $\sigma$ increases the particle/anti-particle energies. The same trend of effects holds true for states with different $(n_r,\ell)$ as shown in Fig.s 3(a), 3(b), 3(c), and 3(d).

Finally. the solutions of some relativistic and non-relativistic quantum mechanical systems, that are of quantum gravity and/or astroparticle physics interest, are often given in terms of the confluent Heun $H_{C}\left( \alpha ,\beta ,\gamma ,\delta ,\eta
,z\right) $ and biconfluent Heun $H_{B}\left( \alpha,\beta ,\gamma ,\delta,z\right)$ functions. The power series expansion of which yields a three terms recursion relations similar to Eq. (\ref{e69}).  The truncation of such power series into polynomials of order $n_r+1$ is a mandatory requirement, so that such solutions are physically admissible, finite and square integrable ones.  For example, the truncation the confluent Heun $H_{C}\left( \alpha ,\beta ,\gamma ,\delta ,\eta ,z\right) $ is secured by the condition $\delta =-\alpha
\left( n_{r}+\left[ \beta +\gamma +2\right] /2\right) $ (c.f., e.g., \cite%
{ref701,ref71}). Ishkhanyan et al. 
\cite{ref71}) have discussed some conditions to be fulfilled for such truncation condition. We, in the current methodical proposal, have introduce yet another alternative condition (i.e.,  $C_{n_{r}+2}=0$, $%
C_{n_{r}+1}\neq 0$, and $C_{n_{r}}\neq 0$ reported for our result in (\ref{e72}) and (\ref{e73})).  that allows such a truncation recipe to be valid/viable and facilitates \textit{conditional exact solvability} of the problem at hand. The same procedure can be followed for the biconfluent Heun $H_{B}\left( \alpha,\beta ,\gamma ,\delta,z\right)$ functions as well. Under such proposal setting (reported in subsection 3-A), one gets \textit{conditional exact solutions}  for a set of quantum mechanical states that are correlated with some condition like that reported in (\ref{e72}).  To facilitate the convergence of the solution into special,  less complicated though, interesting spacetime background models, the reader is advised to follow simple expansion variable that allows such convergence (like the change of variables $y=r^{2}$ used to obtain (\ref{e61})).  Such special spacetime background models work as controlling mechanism on the correctness of the reported solution of the more complicated one. This is documented in subsections 3(b), 3(c), and 3(d).

\end{document}